\documentclass[11pt, executivepaper]{article}
\usepackage[utf8]{inputenc}
\usepackage[T1]{fontenc}
\usepackage{natbib}
\usepackage{amsmath}
\usepackage{mathtools}
\usepackage{xcolor}
\usepackage{amsfonts}
\usepackage{graphicx}
\usepackage{enumitem}
\usepackage{geometry}
\usepackage{hyperref}
\hypersetup{colorlinks= true, allcolors=blue}
\setcitestyle{aysep={}}
\begin{document}

\title{\textbf{Relational Quantum Mechanics and the PBR Theorem: A Peaceful Coexistence}}

\author{Andrea Oldofredi\thanks{Contact Information: Universit\'e de Lausanne, Section de Philosophie, Lausanne, Switzerland. E-mail: Andrea.Oldofredi@unil.ch} \and Claudio Calosi\thanks{Contact Information: Universit\'e de Geneve, Section de Philosophie, Geneve, Switzerland. E-mail: Claudio.Calosi@unige.ch}}

\maketitle

\begin{abstract}
According to Relational Quantum Mechanics (RQM) the wave function $\psi$ is considered neither a concrete physical item evolving in spacetime, nor an object representing the absolute state of a certain quantum system. In this interpretative framework, $\psi$ is defined as a computational device encoding observers' information; hence, RQM offers a somewhat epistemic view of the wave function. This perspective seems to be at odds with the PBR theorem, a formal result excluding that wave functions represent knowledge of an underlying reality described by some ontic state. In this paper we argue that RQM is not affected by the conclusions of PBR's argument; consequently, the alleged inconsistency can be dissolved. To do that, we will thoroughly discuss the very foundations of the PBR theorem, i.e.\ Harrigan and Spekkens' categorization of ontological models, showing that their implicit assumptions about the nature of the ontic state are incompatible with the main tenets of RQM. Then, we will ask whether it is possible to derive a \emph{relational} PBR-type result, answering in the negative. This conclusion shows some limitations of this theorem not yet discussed in the literature.
\vspace{4mm}

\end{abstract}
\vspace{5mm}

\begin{center}
\emph{Accepted for publication in Foundations of Physics}
\end{center}

\clearpage

\tableofcontents
\vspace{5mm}

\section{Introduction}
\label{Intro}

In his seminal essay on Relational Quantum Mechanics (RQM), Carlo Rovelli claims that ``the experimental evidence at the basis of quantum mechanics forces us to accept that distinct observers give different descriptions of the same events'' (\cite{Rovelli:1996}, p.\ 1638). Such a claim, which entails a radically novel perspective on the interpretation of Quantum Mechanics (QM), finds its roots in Einstein's idea behind special relativity, i.e.\ that physical observations depends on the specific reference frame in which interactions occur.\ Furthermore, these insights on the meaning of non-relativistic quantum theory are also motivated by considerations coming from quantum gravity and general relativity, theories where,  according to Rovelli, the relational features of spacetime clearly emerge.

Interestingly, Rovelli claims that RQM is the most natural result of the unification of these ingredients and their application to the formalism of standard quantum theory taken at face value. In the light of this interpretative framework, the notion of an observer-independent state of physical systems is abolished in favor of a perspectival, observer-dependent description of reality (cf.\ \cite{Rovelli:1996, Rovelli:2016, Rovelli:2018}, \cite{Laudisa:2019b}).\ In RQM various observers may provide diverse descriptions of the same sequence of physical events, and such descriptions are equally correct and non-contradictory, \textit{despite their being different}. As a consequence, the values of measurable properties are \emph{relative to} specific observers. 

Rovelli's theory has remarkable foundational implications. We mention a few:

\begin{itemize}
\item A novel meaning of the wave function collapse, simply considered an update of the information possessed by a certain observer relatively to the value of a measured observable. In RQM nothing physical is actually collapsing in measurements interactions, and this is why the tension between the unitary Schr\"odinger dynamics and the collapse postulate is dissolved;
\item A new treatment of Wigner's friend scenarios, both old and new. In RQM the descriptions of a quantum measurement provided by Wigner's friend---who actually measured a certain observable on a quantum system obtaining a definite result---and by Wigner himself---who has no clue about the outcome---are equally correct and non-contradictory (\cite{Frauchiger:2018, Bong:2020});
\item The commitment to a subjective interpretation of quantum probabilities (cf.\ \cite{Smerlak:2007}), without being a subjective version of QM as e.g.\ Quantum Bayesianism, since RQM provides an ontic representation of physical systems, as recently argued by e.g. \cite{Oldofredi:2020b}.
\end{itemize}

This paper focuses on yet another aspect of RQM, namely its interpretation of the wavefunction. According to Rovelli, the wavefunction is considered neither a concrete physical item dynamically evolving in spacetime (or in configuration space), nor an object representing the absolute state of a certain quantum system (cf.\ \cite{Rovelli:1996, Rovelli:2016}). In this context, $\psi$ is defined as a useful computational device encoding the information available to a particular observer about a specific system. Hence, it is generally claimed that RQM offers an epistemic view of the wavefunction. This perspective about the nature of the quantum state, however, seems to be at odds with a formal result obtained by Matthew Pusey, Jonathan Barrett and Terry Rudolph---known as the PBR theorem---according to which all ontic models reproducing the predictions and the statistics of the Born rule must be $\psi$-ontic (cf.\ \cite{PBR:2012}). 
Alternatively stated, as \cite{Leifer:2014} pointed out, such a theorem excludes the possibility that the wavefunction represents the knowledge of an underlying reality described by some \textit{ontic state} (usually denoted $\lambda$). Hence, one would be led to conclude that RQM is in plain contradiction with the PRB theorem.

In this respect, \cite{Laudisa:2019b} claim that
\begin{quote}
[T]he wave function, and more in general the quantum state $\psi$, are interpreted realistically in several presentations of quantum theory. From the perspective of RQM, this is precisely what generates the confusion about quantum theory \citep{Rovelli:2018}. RQM circumvents the theorems for the reality of the wave function (\cite{Leifer:2014}; \cite{PBR:2012}) because it is not a strongly realist theory [...] which is an implicit assumption of these theorems.
\end{quote}

Although we agree that in RQM the wave function is not taken to be a real, concrete object in itself, it should be noted that Rovelli's theory can be properly considered a realist formulation of quantum theory, as argued by several philosophers in the recent years (cf.\ \cite{Dorato:2016}, \cite{Candiotto:2017}, \cite{Oldofredi:2020}, and \cite{Calosi:2020}). Hence, we believe that Laudisa and Rovelli's answer to the issue raised by the PBR theorem can--and should---be developed and refined. Indeed, we think that there is a deeper and more precise reason why RQM and the PRB theorem are not mutually inconsistent. The main aim of the paper is exactly to investigate such a reason. We are going to show why RQM is not affected by the conclusions of PBR's argument and, consequently, this alleged inconsistency can be dissolved.

In order to do this, we will thoroughly discuss the foundations of the PBR theorem, i.e.\ Harrigan and Spekkens' categorization of ontological models (cf.\ \cite{Harrigan:2010}). More precisely, it will be argued that their implicit assumptions about the nature of the ontic state $\lambda$ are incompatible with the basic tenets of RQM. Conforming to this classification, $\lambda$ is considered an observer-independent representation of the state of a certain quantum system. However, the relational character of Rovelli's theory requires that, in order to define correctly what ontic states are, one has to use completely different criteria w.r.t.\ those employed by Harrigan and Spekkens. According to the main tenets of RQM, in fact, \emph{both} $\psi$ and $\lambda$ must be relational, meaning that 
\begin{itemize}
\item  $\lambda$ represents the ontic state of quantum systems \textit{relatively} to a certain observer,
\item $\psi$ stores information that a particular observer has \textit{relatively} to a given system.
\end{itemize}
\noindent Thus, in the following sections we will carefully explain which assumptions RQM makes about $\lambda$, and how they diverge from those employed by Harrigan and Spekkens. In particular, it will be argued that Harrigan and Spekkens' approach does not have the necessary formal and ontological resources to be correctly applied to RQM. This has a remarkable consequence---or so we contend: given that PBR theorem relies on such a classification of quantum models, but the latter fails to capture some basic tenets of RQM, one can safely conclude that Rovelli's theory does not lie within the scope of the theorem, avoiding any formal contradiction with it. In addition, we will ask whether it is possible to derive a PBR-type result in the context of RQM. We will answer in the negative. This conclusion also shows some limitations of the PBR-theorem that, to the best of our knowledge, have not been discussed in literature.

The structure of the paper is the following: in Section \ref{Stage} we introduce the essential elements of our discussion, providing a brief description of RQM, Harrigan and Spekkens' categorization of ontological models, and the PBR theorem (readers familiar with the literature may skip this part, though our reconstruction of the PBR-theorem is somewhat unorthodox). In Section \ref{RQM_PBR} we will argue that RQM and the PRB theorem can peacefully coexist, whereas in Section \ref{RPBR} a relational PBR-type result will be discussed taking into account several scenarios involving different observers. Section \ref{conc} concludes the paper. 

\section{Setting the Stage: RQM and the PBR Theorem}
\label{Stage}

\subsection{RQM in a Nutshell}
\label{2.1}

RQM was first proposed in \cite{Rovelli:1996}. In its original formulation it consisted of two parts:

\begin{itemize}
\item A re-interpretation of the usual quantum formalism;
\item A derivation of this formalism from basic, general principles.
\end{itemize}

The focus of this paper is restricted to the first part. We follow closely \cite{Rovelli:1996} here. Rovelli explicitly understands RQM as a ``generalization'' of the standard Copenhagen interpretation of quantum mechanics. This generalization is twofold. First, there is no privileged notion of an ``observer''---thus \textit{every quantum system is on a par}:

\begin{quote}
[R]elational QM is Copenhagen quantum mechanics made democratic by bringing all systems onto the same footing (\cite{Rovelli:2018}, p.\ 11).
\end{quote}

\noindent Thus, the notion of an ``observer'' does not

\begin{quote}
[M]ake any reference to conscious, animate, or computing, or in any other manner \textit{special}, system (\cite{Rovelli:1996}, p.\ 1641).
\end{quote}

\noindent In the light of this, any physical system can count as an ``observer''. Second, there is no privileged notion of ``measurement''---thus \textit{every physical interaction is on a par}:

\begin{quote}
[M]easurement is an interaction like any other (\cite{Rovelli:2018}, p.\ 5).
\end{quote}

There are four ``basic tenets'' of the Copenaghen interpretation that, when suitably adjusted, RQM is supposed to retain:

\begin{description}
\item \textbf{Eigenfunction-Eigenvalue Link} (EEL). A physical system $s$ has a definite value $v$ of an observable $O$ iff the state of $s$ is an eigenstate of $O$ that belongs to $v$.

\item \textbf{Schr\"{o}dinger Dynamics}. The state of $s$ evolves according to the Schr\"odinger equation 
\begin{align}
H(t) |\psi(x,t)\rangle = i\hbar\frac{\partial}{\partial t} |\psi(x,t)\rangle.
\end{align}

\item \textbf{``Collapse'' Postulate}. At the time of measurement the state of $s$ collapses into one of the eigenstates of $O$ with probability given by the Born rule.

\item  \textbf{Born rule}. If an observable quantity $O$ is measured:
\begin{itemize}
\item the result will be one of the eigenvalues $o_i$ of $O$;
\item the probability $P$ to obtain eigenvalue $o_i$ is given by $\langle\psi|P_i|\psi\rangle$.
\end{itemize}  
\end{description}

As we will see in due course, caution is due to interpret the \textbf{Collapse Postulate}. As of now, we think RQM is best appreciated by focusing on what Rovelli calls ``The Third Person'' (or ``third system'') problem. Consider an observable $O$ with two eigenvectors $|+\rangle$ and $|-\rangle$, with eigenvalues $+1$ and $-1$ respectively.  Suppose we have a physical system $s_1$, that, at time $t_1$, is in a superposition of $O$-states:

\begin{equation}
|\psi\rangle_{s_1} =  (c_1 |+\rangle + c_2|-\rangle)_{s_1}
\end{equation}

A second system $s_2$ interacts with $s_1$ in what the Copenaghen interpretation will call a ``measurement'' of $O$, and finds $s_1$ to have value $O = + 1$. According to the \textbf{``Collapse'' Postulate''} we have:

\begin{gather}
\label{2}
t_1 \rightarrow t_2  \\ 
\nonumber
(c_1 |+\rangle + c_2 |-\rangle)_{s_1} \rightarrow  |+\rangle_{s_1}
\end{gather}

\noindent The result of the quantum interaction between $s_1$ and $s_2$ is that $s_1$ \textit{acquires} a definite value property, namely $O = +1$---this much follows from EEL. Consider now a system $s_3$ that does not interact with either $s_1$ or $s_2$. Its description of the quantum situation encoded in \eqref{2} will only refer to the \textbf{Schr\"odinger Dynamics}. Thus, we get:

\begin{gather}
\label{3}
t_1 \rightarrow t_2  \\ 
\nonumber
((c_1 |+\rangle + c_2 |-\rangle \otimes |init\rangle)_{s_{12}} \rightarrow (c_1|+\rangle  \otimes |plus\rangle + c_2 |-\rangle \otimes |minus\rangle)_{s_{12}}
\end{gather}

\noindent where, as conventions, $s_{12}$ is the quantum system comprising $s_1$ and $s_2$ and states $|+\rangle$ and $|plus\rangle$ are correlated in the obvious way.\footnote{Clearly, the same goes for $|-\rangle$ and $|minus\rangle$.}

It is immediately clear that the descriptions of the very same events given by $s_2$ and $s_3$ are very different: according to $s_2$, $s_1$ is in an eigenstate of $O$. Thus, $s_1$ has the definite value property $O= +1$. This is not the case for $s_3$. According to $s_3$, $s_1$ is not in an eigenstate of $O$, and therefore does not have any definite value property of $O$.\footnote{Different metaphysical readings of this situation have been proposed in the literature.\ \cite{Dorato:2016} argues in favor of a dispositionalist account, whereas \cite{Calosi:2020} argue for an indeterminate account.} This is what Rovelli calls \textbf{Main Observation}:

\begin{description}
\item \textbf{Main Observation}. In QM different systems may give different accounts of the same sequence of events.
\end{description}

RQM can be understood as that interpretation of quantum mechanics that, by contrast with more familiar ones, holds that \textit{both} descriptions \eqref{2} and \eqref{3} are correct. But, how can that be? These descriptions are clearly different, as we saw already. One possibility is simply that, both are correct, and yet one (or both) is only a partial description description of quantum phenomena. Rovelli rejects such a possibility explicitly. Quantum formalism, at least in its usual applications, is complete:

\begin{description}
\item \textbf{Completeness}. QM provides a complete and self-consistent scheme of description of the physical world.
\end{description}

There seems to be one other possibility. Descriptions \eqref{2} and \eqref{3} are \textit{different}, \textit{correct}, and \textit{complete} descriptions of the relevant quantum events because they are correct \textit{relatively to different relativization targets}. That is to say, description \eqref{2} is a correct and complete description of quantum events at $t_1 \rightarrow t_2$ \textit{relative} to $s_2$, whereas description \eqref{3} is a correct and complete description of the \textit{very same} quantum events \textit{relative} to $s_3$. This is in effect \textit{the} basic tenet of RQM:

\begin{description}
\item \textbf{Basic Tenet of RQM}. RQM dictates a \textit{relativization of states and observables of physical systems to other physical systems}.
\end{description}

\noindent In Rovelli's own words:

\begin{quote}
[Q]uantum mechanics is a theory about the physical description of \textit{physical systems relative to other systems}, and this is a complete description of the world (\cite{Rovelli:1996}, p.\ 1650).
\end{quote}

\begin{quote}
[T]he actual value of \textit{all} physical quantities of \textit{any} system is only meaningful in relation to another system (\cite{Rovelli:2018}, p.\ 6, italics in the original).
\end{quote}

We could at this point, slightly change the formalism in order to incorporate relativization explicitly in the formalism itself. For instance we could write $ O= v_1 (s_i/s_j)$ for ``$s_i$ has value $v_1$ of observable $O$ relative to $s_j$''. More importantly for our purposes, we suggest to write:

\begin{equation}
|\psi\rangle_{s_i/s_j}
\end{equation}

\noindent for ``$s_1$ is in state $\psi$ relative to $s_j$'', and

\begin{equation}
\lambda_{s_i/s_j}
\end{equation}

\noindent for ``$s_i$ has an ontic state $\lambda$ relative to $s_j$''---the rationale behind this direct incorporation in the formalism will be clear in due course. If so, the quantum phenomena we started from, i.e., the ones in the ``Third Person Problem'', will be more perspicuously represented---at $t_2$---by:

\begin{gather}
|\psi\rangle_{s_1/s_2} = |+\rangle_{s_1} \\ 
\nonumber
|\psi\rangle_{s_{12}/s_3} = (c_1|+\rangle  \otimes |plus\rangle + c_2 |-\rangle \otimes |minus\rangle)_{s_{12}}
\end{gather} 

Before we move on to the next section we should pause to consider a crucial issue. Rovelli himself presents RQM using quantum states, and frequently uses talk of ``relativization of states''. This clearly does not entail that he recognizes a substantive ontological reading of such quantum states. In effect, Rovelli is explicitly skeptical about such an heavy-weight ontological reading. He writes: 

\begin{quote}
[T]he \textit{conceptual} step was to introduce the notion of ``wavefunction'' $\psi$, soon to be evolved in the notion of ``quantum state'' $\psi$, endowing it with heavy ontological weight. This conceptual step was \textit{wrong, and dramatically misleading}. We are still paying the price for the confusion it generated (\cite{Rovelli:2018}, p.\ 2, italics added). 
\end{quote}

This will play a crucial role. As of now, it is unclear what the (alleged) ``heavy ontological weight'' of the quantum state amounts to.\ \cite{Harrigan:2010} provides a clear framework to define rigorously such a notion by classifying them either as \textit{ontic} or \textit{epistemic}. To this we now turn.

\subsection{Harrigan and Spekkens' Distinction: $\psi$-Ontic and $\psi$-Epistemic Models}
\label{2.2}

In order to illustrate the distinction between ontic and epistemic states, let us consider a straightforward example of a classical particle (cf.\ \cite{Leifer:2014}). In classical mechanics one assigns a precise position $q$ and a momentum $p$ to it, and the pair $(q, p)$ constitutes the phase space point of the object under consideration. This point describes the physical state of the particle, i.e.\ its ontic state, which exists objectively and independently of any observer.\footnote{Its future positions and momenta, moreover, are univocally determined via Hamilton's equations, meaning that at any time the ontic state of the particle will be described by $(q, p)$. In addition, every other observable property of the relevant physical system is a function of $(q, p)$.}\ However, if one does not have precise information about the position and momentum of this physical system, then one can represent the uncertain knowledge about its ontic state by a probability density $f(q, p)$ over phase space. Interestingly, $f(q, p)$---the epistemic state---represents mere knowledge, and does not describe any inherent property of the particle.\ Different observers may not have equal knowledge of the state of our particle. If so, it is possible to associate several epistemic states to the same ontic state. In effect, such epistemic states---i.e.\ such probability distributions---may overlap in the particle's phase space.\ By contrast, the particle's ontic state is uniquely determined by the point $(q, p)$. 

Harrigan and Spekkens' categorization of ontological quantum models employs the ontic/epistemic distinction in order to determine whether the quantum state $\psi$ has to be interpreted as representing some underlying reality or just observers' knowledge of particular systems (cf.\ \cite{Harrigan:2010}). This constitutes the very theoretical heart of the PBR theorem. The taxonomy is framed within operational quantum theory, an approach to QM where primitive notions consist exclusively in preparations procedures, i.e.\ instructions about how to prepare physical systems in certain states, and measurements performed on such states. Harrigan and Spekkens claim that 

\begin{quote}
[I]n an operational formulation of quantum theory, every preparation $P$ is associated with a density operator $\rho$ on Hilbert space, and every measurement $M$ is associated with a positive operator valued measure (POVM) $\{ E_k\}$. (In special cases, these may be associated with vectors in Hilbert space and Hermitian operators respectively.) The probability of obtaining outcome $k$ is given by the generalized Born rule, $p(k|M, P)=\textrm{Tr}(\rho E_k)$ (\cite{Harrigan:2010}, p.\ 128).
\end{quote}

\noindent The main aim of such operational models, they continue, is to provide probabilities $p(k|M, P)$ of outcomes $k$ for some measurement $M$ performed on prepared systems, given a set of preparation instructions $P$.\ When a measurement is carried out, a measuring device will ``reveal something about those properties'' (\emph{ibid.}). In this respect, Harrigan and Spekkens assume that a complete specification of the properties of a given individual physical system under scrutiny is provided by its ontic state, denoted by $\lambda$.\footnote{As in the classical case discussed above, the ontic state $\lambda$ belongs to an ontic state space $\Lambda$.} More precisely, $\lambda$ provides a complete specification of the preparation procedures that are performed on a particular quantum system, and hence, it yields a complete description of its measurable properties.\footnote{It is worth noting, however, Harrigan and Spekkens use the term ``ontological'' in a much weaker sense with respect to standard philosophical jargon: they simply mean that $\lambda$ refers to something real in the world, leaving many details regarding the metaphysics of quantum objects completely unspecified.} Furthermore, they underline that although an observer knows exactly the preparation procedures prior the performance of a certain measurement, she may not have complete knowledge about the ontic state $\lambda$ of the system under investigation. Thus, it follows that while the future outcomes $k$ of a certain measurement are determined by $\lambda$---and therefore the probability to obtain them is given by $p(k|\lambda, M)$---the epistemic state of the experimenter is represented only by $p(\lambda| P)$. Notably, an observer that has incomplete information about $\lambda$ assigns ``non-sharp'' probability distributions over the ontic state space $\Lambda$, i.e.\ multiple probability distributions can be assigned to the very same ontic state, exactly as in the classical case discussed a few lines above. 

Against this background, Harrigan and Spekkens go on  to provide conditions to classify ontological quantum models as $\psi-$ontic or $\psi-$epistemic. Briefly and roughly, a model is defined $\psi-$ontic if the ontic state $\lambda$ can be consistently described by a \textit{unique} pure state. Consequently, in $\psi-$ontic models different quantum states correspond to disjoint probability distributions over the space of ontic states $\Lambda$. More precisely, a model is said to be $\psi-$ontic if ``for any pair of preparation procedures, $P_{\psi}$ and $P_{\phi}$, associated with distinct quantum states $\psi$ and $\phi$, we have $p(\lambda | P_{\psi})p(\lambda|P_{\phi})=0$ for all $\lambda$'' (\cite{Harrigan:2010}, pp.\ 131-132). That is, observers' epistemic states associated with different quantum states do not overlap in $\Lambda$. Conversely, a model is defined $\psi-$epistemic if there exists ontic states consistent with \textit{more than one pure state}; in such epistemic models, thus, there are quantum states that correspond to overlapping probability distributions in $\Lambda$. In this operational context, this implies that agents' epistemic states may overlap, i.e.\ there exist preparation procedures $P_{\psi}, P_{\phi}$ such that $p(\lambda | P_{\psi})p(\lambda|P_{\phi})\neq0$. Hence, the ontic state $\lambda$ can be consistently represented by both quantum states $\psi$ and $\phi$: 

\begin{quote}
[I]n a $\psi-$epistemic model, multiple distinct quantum states are consistent with the same state of reality---the ontic state $\lambda$ does not encode $\psi$'' (\emph{ibid.}, p.\ 132). 
\end{quote}

In the case of $\psi-$epistemic models, then, the quantum state refers to observers' incomplete knowledge of reality, it is not a description of reality itself.

On top of that, Harrigan and Spekkens divide quantum models in $\psi-$\emph{complete} and $\psi-$\emph{incomplete}. In $\psi-$complete models, the pure quantum state encodes every information about the represented physical system. Moreover, in such models there is a one-to-one relation between reality and its complete description provided by $\psi$. If one knows the pure quantum state of a certain system under consideration, one has a complete knowledge of its ontic state. Therefore, $\psi-$complete models are also $\psi-$ontic. Examples of such models are given by standard QM, Everett's relative-state formulation (cf.\ \cite{Everett:1957aa}), the Many-World interpretation (cf.\ \cite{Wallace:2012aa}), and Wave-Function Realism (cf.\ \cite{Albert:2013}). 

If $\psi$ does not represent reality completely, then a quantum model is $\psi-$incomplete. Notably, $\psi-$incomplete models may be either $\psi-$supplemented or $\psi-$epistemic. In the former case, the description of a physical system provided by $\psi$ is supplemented by some additional (or hidden) variables, whose value is generally unknown. In hidden variables models, trivially, the quantum state provides partial or incomplete knowledge of the system. Well-known examples of hidden variables models are Bohmian mechanics (cf.\ \cite{Durr:2013aa}), Bohm's pilot-wave theory (cf.\ \cite{Bohm:1952aa}), and Nelson's stochastic mechanics (cf.\ \cite{Nelson:1966aa}). Notably, also the class of $\psi-$supplemented models is $\psi-$ontic. Finally, in the case of $\psi-$epistemic models, $\psi$ represents agents' incomplete knowledge of reality, and not reality itself. Typical examples of such a kind of models is given by QBism (cf.\ \cite{Fuchs:2002}) and RQM.\footnote{From these distinctions some conclusions can be derived. Firstly, there cannot be models which are simultaneously $\psi-$complete \emph{and} $\psi-$epistemic. Thus, if a model is $\psi-$complete, it must be $\psi-$ontic (cf.\ Lemma 6, \cite{Harrigan:2010}, p.\ 133). Conversely, if a model is $\psi-$epistemic, it cannot be $\psi-$ontic, since it describes only observers' knowledge and not any underlying physical reality. Secondly, if a model is $\psi-$incomplete, then it can either be $\psi-$ontic, as in the case of hidden variable models, or $\psi-$epistemic.}

In the rest of the paper we will critically analyze some implicit assumptions in Harrigan and Spekkens' categorization. In particular we will argue that some requirements concerning the nature of $\lambda$ are in tension, if not inconsistent, with basic ontological tenets of RQM. As we pointed out already, this will have notable consequences for the relation between RQM and the PBR theorem, a formal result that we are now going to introduce.

\subsection{The PBR Theorem}
\label{2.3}

The PBR theorem was heralded as the single most important result in quantum foundations after the Bell-theorem.\footnote{This judgement was given by Anthony Valentini, cf.\ \cite{Reich:2011}.} It allegedly establishes that an epistemic reading of the quantum state cannot recover quantum predictions. Pusey, Barrett and Rudolph phrase the result in the form of a no-go theorem:

\begin{quote}
[T]his Article presents a no-go theorem: if the quantum state merely represents information about the real physical state of a system, then experimental predictions are obtained that contradict those of quantum theory (\cite{PBR:2012}, p.\ 475).
\end{quote}

\noindent We present here a streamlined reconstruction of the result. The argument depends on different assumptions, but we will mention just a few:

\begin{description}

\item \textbf{Preparation Independence}. It is possible to prepare different physical systems in such a way that their ontic states are independent, i.e. uncorrelated.

\item \textbf{Harrigans and Spekkens Definition of Epistemic State}. Quantum states $|\psi\rangle$ and $|\phi\rangle$ associated with preparation procedures $P_{\psi}$ and $P_{\phi}$ are \textit{epistemic states} iff $p(\lambda | P_{\psi})p(\lambda|P_{\phi})\neq0$.\footnote{We saw in the previous subsection this is really not the definition of an epistemic state, but rather it is implied by the definition. The reason we use this will be clear in a  minute.}

\item \textbf{Measurement Response}. If two quantum systems are prepared in such a way as to fulfill \textbf{Preparation Independence}, measurements respond solely to the properties of the system that is being measured.
\end{description}

We already discussed \textbf{Harrigans and Spekkens Definition of Epistemic State}. As for \textbf{Measurement Response}, let us quote directly from the original PBR-paper:

\begin{quote}
[T]he outcome of the measurement can only depend on the physical states of the two systems at the time of measurement (\cite{PBR:2012}, p.\ 476).
\end{quote}

\noindent As we will see, this will play a crucial role for our discussion. As of now, let us go back to proof of the PBR theorem. \\

Let $s_1$ be a quantum system prepared with two different procedures $P_\psi$ and $P_\phi$. Let $\langle \psi|\phi\rangle = \frac{1}{\sqrt{2}}$. Consider a two-dimensional Hilbert space $\mathcal{H}$ spanned by $|\psi\rangle = |\uparrow\rangle$ and $|\phi\rangle = |+\rangle = \frac{1}{\sqrt{2}}(|\uparrow\rangle + |\downarrow\rangle)$. Suppose now $|\uparrow\rangle$ and $|+\rangle$ are epistemic states. Then the ontic state of $s_1$ should be compatible with both and we can write:
\begin{align}
\label{8}
p(\lambda_1|P_{\uparrow}) \neq 0,\\ \nonumber
p(\lambda_1|P_{+}) \neq 0.
\end{align}

Prepare another system $s_2$ in such  a way that the ontic states of the two systems are uncorrelated---as guaranteed by \textbf{Preparation Independence}. Then, by the same argument:
\begin{align}
\label{9}
p(\lambda_2|P_{\uparrow}) \neq 0,\\ \nonumber
p(\lambda_2|P_{+}) \neq 0.
\end{align}

Equations \eqref{8} and \eqref{9} are meant to capture that the  ontic states of $s_1$ and $s_2$, $\lambda_1$ and $\lambda_2$ respectively, are compatible with both $|\uparrow\rangle$ and $|+\rangle$. Then, the complex system $s_{12}$ is compatible with the following tensor product states:
\begin{align}
\label{10}
|\omega_1\rangle_{12} = |\uparrow\rangle_1 \otimes |\uparrow\rangle_2 \\ \nonumber
|\omega_2\rangle_{12} = |\uparrow\rangle_1 \otimes |+\rangle_2 \\ \nonumber
|\omega_3\rangle_{12} = |+\rangle_1 \otimes |\uparrow\rangle_2 \\ \nonumber
|\omega_4\rangle_{12} = |+\rangle_1 \otimes |+\rangle_2 \\ \nonumber
\end{align}

If the ontic state $\lambda_{12}$ is compatible with all of them, we can write:
\begin{align}
\label{11}
p(\lambda_{12}|P_{\omega_1}) \neq 0 \\ \nonumber
p(\lambda_{12}|P_{\omega_2}) \neq 0 \\ \nonumber
p(\lambda_{12}|P_{\omega_3}) \neq 0 \\ \nonumber
p(\lambda_{12}|P_{\omega_4}) \neq 0 \\ \nonumber
\end{align}

Set $|-\rangle= \frac{1}{\sqrt{2}} (|\uparrow\rangle - |\downarrow\rangle)$. Now we perform a measurement on $s_{12}$ that projects onto the following states:
\begin{align}
\label{12}
|\chi_1\rangle_{12} = \frac{1}{\sqrt{2}}(|\uparrow\rangle_1 \otimes |\downarrow\rangle_2 + |\downarrow\rangle_1 \otimes |\uparrow\rangle_2) \\ \nonumber
|\chi_2\rangle_{12} = \frac{1}{\sqrt{2}}(|\uparrow\rangle_1 \otimes |-\rangle_2 + |\downarrow\rangle_1 \otimes |+\rangle_2) \\ \nonumber
|\chi_3\rangle_{12} = \frac{1}{\sqrt{2}}(|+\rangle_1 \otimes |\downarrow\rangle_2 + |-\rangle_1 \otimes |\uparrow\rangle_2) \\ \nonumber
|\chi_4\rangle_{12} = \frac{1}{\sqrt{2}}(|+\rangle_1 \otimes |-\rangle_2 + |-\rangle_1 \otimes |+\rangle_2) \\ \nonumber
\end{align}

\noindent This measurement responds solely to the properties of $s_{12}$, given \textbf{Measurement Response}. Now, algebraic manipulation shows that
\begin{equation}
\langle \omega_1 | \chi_1\rangle = \langle \omega_2 | \chi_2\rangle = \langle \omega_3 | \chi_3\rangle = \langle \omega_4 | \chi_4\rangle = 0
\end{equation}

That is to say that for every measurement direction in \eqref{12} we can find a state of $s_{12}$ in \eqref{10} that is orthogonal to it. In this case, standard quantum predictions yield that:
\begin{align}
\label{14}
p(\lambda_{12}|P_{\omega_1}) = 0 \\ \nonumber
p(\lambda_{12}|P_{\omega_2}) = 0 \\ \nonumber
p(\lambda_{12}|P_{\omega_3}) = 0 \\ \nonumber
p(\lambda_{12}|P_{\omega_4}) = 0 \\ \nonumber
\end{align}
\noindent contradicting \eqref{11}. Thus quantum states are not epistemic states. Or so the argument goes.

\section{The Peaceful Coexistence}
\label{RQM_PBR}

Given the formal result above, it seems to follow that RQM contradicts the predictions of standard quantum mechanics. The argument seems straightforward: according to Rovelli, $\psi$ is to be interpreted epistemically.  And the PBR theorem (allegedly) establishes that an epistemic interpretation of quantum states cannot reproduce standard quantum predictions. However, it is well-known that RQM is in perfect empirical agreement with respect to the predictions and statistics of quantum theory, the PBR argument notwithstanding. How can this be? How can we dissolve the tension between these two apparently incoherent claims? 
In what follows we answer to these questions. \footnote{To be clear: it is not the aim of the paper to provide a thorough defense of RQM. Rather, it is to show that there the PBR-theorem does not proivde an argument against it.}
\vspace{2mm}

In order to show how RQM can escape the consequences of PBR theorem, we need to focus on the implicit assumptions made by Harrigan and Spekkens concerning the nature of the ontic state $\lambda$.\ In Section \ref{2.2} we saw that ontological models are framed within operational quantum theory, and that their basic ingredients are (i) properties of individual physical systems associated with experimental procedures, and (ii) measurements performed on such systems revealing the value of some such particular properties. In effect, Harrigan and Spekkens assume that the ontic state $\lambda$ of a quantum system provides a \emph{complete} description of its \emph{inherent} properties (\cite{Harrigan:2010}, p.\ 128)---i.e.\ attributes that objects instantiate in virtue of the way themselves are, independently of any relation they  stand in with external observers and/or contexts (cf.\ \cite{Lewis:1986aa}, p.\ 61). Moreover, $\lambda$ provides an \emph{objective} representation of quantum systems, since the ontic state of a quantum object refers to an underlying physical reality whose features do not depend upon any particular observer. Hence, according to Harrigan and Spekkens, we can uniquely characterize a certain system through the specification of its instantiated properties by specifying a list of preparation procedures.\ In the light of this, $\lambda$ is taken to be an observer-independent, complete description of physical systems.\footnote{This is in line with the analysis in e.g.\  \cite{Oldofredi:2020b}.} 

For these reasons, in $\psi$-complete models---as e.g.\ standard QM---the authors claim that:
\begin{itemize}
\item variations of $\lambda$ cause variations in the wave function $\psi$;
\item there is a one-to-one correspondence (a bijective map) between ontic states of physical systems and pure quantum states representing them, since there is a \textit{unique correct description} of the various quantum systems under consideration. 
\end{itemize}

\noindent From these premises, it follows straightforwardly that if a certain $\lambda$ can be described by multiple $\psi$-s, then these $\psi$s are taken to represent the incomplete knowledge of the ontic state possessed by different observers. Hence, it is easy to understand why overlapping probability distributions over $\Lambda$ are defined as epistemic states: these represent the lack of information (ignorance) of a certain observer about the complete specification of the properties of a given system. Thus, if multiple wave functions can be ascribed to the same $\lambda$, they equally fail to describe some property in the ontic state of the object at hand. 

These assumptions about the nature of $\lambda$ are clearly violated in the context of RQM. Recall the Third Person scenario presented in Section \ref{2.1}. In that case, it is clear upon inspection that the ontic state of a particular system strictly depends on the perspective of a given observer, since physical systems are represented by relational states. In effect, according to the observer $s_2$, the system $s_1$ is in the $|+\rangle$ state, and thus, it has a definite value for the observable $O$. By contrast, according to $s_3$, $s_1$ system is in a superposition of states, and hence it does not possess any definite value for $O$. However, from the \textbf{Main Observation}, we know that in RQM both $s_2$ and $s_3$ provide \textit{correct} descriptions of a given sequence of physical events. This immediately entails that in RQM the ontic state of physical systems (and thereby their properties) depends on the perspective of a specific observer, in plain contrast with Harrigan and Spekkens' entire farmework.

Now, if the relational nature of quantum systems entails that their ontic states depend on the specification of a certain perspective, then different observers can ascribe diverse wave functions to the very same $\lambda$. Against this background, however, these $\psi$s \textit{do not represent epistemic limitations}, such as, for instance, incomplete knowledge. Rather, in RQM \textit{they simply reflect the perspectival nature of physical systems}. Alternatively, given that physical systems in RQM are described by relational properties, which by definition require the specification of a certain observer and/or context to be meaningful, different observers may assign different ontic states to the same target object, as in the already mentioned example of the Third Person problem---and as we are about to see again in a different context. Such relational states, however, should be not interpreted epistemically, but ontologically: they reflect that reality (and thus the ontic states of quantum objects) is observer-dependent, or better, perspectival, contrary to what Harrigan and Spekkens simply assume. 
It is thus clear that RQM employs completely different criteria to define what ontic states are with respect to those assumed by Harrigan and Spekkens. In particular, according to Rovelli's theory it is possible to assign several $\psi$s to the same observed system without necessarily be committed to a $\psi$-epistemic view as defined by their classification. This entails that the wavefunction must be relational as well: $\psi$ can no longer be associated with a perspective-independent ontic state, but rather with a relational $\lambda$, i.e.\ an underlying perspectival reality. Thus, we conclude, according to RQM,  \emph{both} $\psi$ and $\lambda$ must be relational, that is: 
\begin{itemize}
\item  $\lambda$ represents a quantum system always \textit{relatively} to a certain observer,
\item $\psi$ stores information that a particular observer has \textit{relatively} to a given system.
\end{itemize}

\noindent To sum up: given that in RQM an absolute representation of physical systems is unavailable by construction, there cannot be a one-to-one correspondence between $\lambda$ and $\psi$. Strictly speaking, there is neither an \textit{absolute}, i.e.\ observer-independent, $\lambda$, nor an \textit{absolute} $\psi$.
RQM allows to assign more than one $\psi$ to the very same individual physical system. Remarkably, such $\psi$s are not interpreted as a manifestation of the ignorance of the various observers, but are a simple consequence of the relational ontology of the theory. As a matter of fact, one could see some irony here. Suppose we were to insist that there \textit{must} be a one-to-one correspondence between $\lambda$ and  $\psi$. Then, given that different observers assign different $\lambda$-s to the same quantum systems, they do assign different $\psi$-s as well. 

Even if RQM supports a somewhat epistemic take on the quantum wavefunction, \textit{this differs substantially from Harrigan and Spekkens definition of $\psi$-epistemic models}. In particular, although in RQM it is possible to ascribe multiple wave functions to the very same system, it is not the case that these descriptions fail to capture certain intrinsic, observer-independent properties instantiated by a certain physical system. There are no such properties to be captured to begin with! Therefore, since the assumptions employed to define what ontic states are in RQM differ completely with respect to those at play in \cite{Harrigan:2010}, we conclude that this classification does not have the correct metaphysical and formal resources to deal with relational ontic states and relational $\psi$s. Thus, Harrigan and Spekkens' classification cannot straightforwardly be applied to RQM. This fact has a remarkable implication for our discussion: given that the PBR theorem crucially relies on Harrigan and Spekkens' classification of quantum ontological models, as per the \textbf{Harrigans and Spekkens Definition of Epistemic States} assumption, but the latter cannot be used to evaluate RQM, one can safely conclude that RQM does not lie within the scope of the theorem. This means that when Pusey, Barrett and Rudolph refer to $\psi$-epistemic model, they are not referring to RQM. There is no contradiction between the epistemic interpretation of $\psi$ in RQM and the PBR result.  

From such a conclusion, we can also infer a more general lesson, i.e.\ that it is not necessarily true that the PBR theorem excludes every type of $\psi$-epistemic model, or better, not every $\psi$-epistemic model contradicts the predictions and statistics of standard QM, \emph{contra} the main claim of the PBR argument.

\section{A Relational PBR Theorem?}
\label{RPBR}

In the light of the above, a natural question arises: could we find a \textit{relational counterpart} of the PBR-theorem?\ That is, could we give an analogous result once the relativization of states, observables and ontic states characterizing RQM is taken explicitly into account? In this section we provide an argument that answers the question in the negative.

From our reconstruction, the PBR theorem crucially depends on three assumptions. Our argument in Section \ref{RQM_PBR} can be read as a challenge to the second, namely to the \textbf{Harrigan and Spekkens Definition of Epistemic State} assumption. In particular, it can be read as challenging the \textit{exhaustiveness} of Harrigan and Spekkens's classification. This is because we argued that their categorization importantly leaves out RQM---or more precisely, that it cannot be consistently used to evaluate RQM.\footnote{This assumption can be challenged on other grounds as well, as done for instance in \cite{Ladyman:2021}. Hance et al. argue  that wavefunctions can represent simultaneously reality \emph{and} knowledge, contrary to Harrigan and Spekkens definitions which are mutually exclusive. Moreover, the paper shows on the one hand that such exclusiveness is simply presupposed, and on the other that it is neither necessary, nor sufficient to capture the relation existing between the ontic state $\lambda$ and $\psi$.}
The \textbf{Preparation independence} assumption has been the focus of a great deal of scrutiny as well.\footnote{For critical discussions about the validity of this premise cf.\ \cite{Mansfield:2016}, \cite{Schlosshauer:2012, Schlosshauer:2013, Schlosshauer:2014} and references therein. Interestingly, dropping this assumption altogether, \cite{Lewis:2012} showed with an explicit model that it is mathematically possible to interpret quantum states as agent's information concerning an underlying physical state. In addition, \cite{Mansfield:2014} shows that the PBR theorem ``breaks down'' in models that relax \textbf{Preparation independence}.} To our knowledge, however, \textbf{Measurement Response} has not yet been properly discussed. We now turn to argue that RQM sheds interesting new light on this assumption as well. In particular, as we are going to show, it seems that its use in the context of RQM undermines one crucial step in the derivation of the PBR-theorem.\\

As we saw in Section \ref{2.1}, according to the main tenet of RQM, states and observables must always be relativized to other physical systems functioning as observers. We suggested to incorporate this relativization directly into the formalism; going back to the inner workings of the PBR-theorem this results in the following.\ An observer $s$, which is a physical system like any other, prepares two quantum systems $s_1$ and $s_2$ which can be both compatible with the epistemic states $|\psi\rangle$ and $|+\rangle$ encountered in Section \ref{2.3}, implementing two different protocols $P_{\psi}$ and $P_{\phi}$ and using two copies of the same preparation device.\ We have to recall here that in RQM every quantum state is epistemic, since it refers to the knowledge that a given observer has about a certain system. Moreover, although $s$ knows the instructions to prepare the systems $s_1, s_2$, the agent ``may nonetheless have incomplete knowledge of $\lambda$'' (\cite{Harrigan:2010}, p.\ 128). As a consequence of RQM, after the preparations protocols have been implemented, $s_1$ and $s_2$ acquire a particular ontic state \textit{relative} to $s$, i.e.\ $\lambda_{1/s}$, and $\lambda_{2/s}$ respectively---where, for the sake of simplicity, we \textit{assumed} that relativization to the same physical system is consistent.\footnote{To this regard, we have reasons to believe that our assumption is consistent. Indeed, discussing the third person scenario we saw a similar situation; an observer $s_2$ interacted with a system $s_1$ which before the interaction was in a superposition of $O$-states, and found $s_1$ to have a definite value for the observable $O$. In the case discussed in the present section, an observer $s$ prepares two systems interacting directly with them, thus, they will acquire a certain state relative to $s$. Here we make a standard maneuver in operational quantum mechanics, namely to consider preparations as measurements.} Hence, equations \eqref{8} and \eqref{9} should be replaced with their relational counterparts: 

\begin{align}
\label{15}
p(\lambda_{1/s}|P_{\uparrow/s}) \neq 0,\\ \nonumber
p(\lambda_{1/s}|P_{+/s}) \neq 0
\end{align}
\noindent and
\begin{align}
\label{16}
p(\lambda_{2/s}|P_{\uparrow/s}) \neq 0,\\ \nonumber
p(\lambda_{2/s}|P_{+/s}) \neq 0.
\end{align}

\noindent Interestingly, in this scenario \textbf{Preparation Independence} is trivially met since both systems are prepared at the same site by the same agent $s$ who can freely choose among the possible protocols how to prepare $s_1$ and $s_2$. After their preparation, suppose that the two systems are then brought together to form a complex system $s_{12}$ as required in the proof of the PBR theorem. Referring to this, it is worth noting that \cite{PBR:2012} implicitly assume that the composite system $s_{12}$ is simply given by the product of the independent physical states of the individual systems forming it. While this not a mereologically trivial fact, we accept it for the sake of the argument; then, the ontic state of the complex system $s_{12}$ relative to the agent $s$ is given by $\lambda_{12/s}$.\footnote{Cf.\ \cite{Schlosshauer:2013} for critical remarks concerning this mereological issue.} 
Moreover, let us suppose that the ontic state $\lambda_{12/s}$ is compatible with all possible tensor product states of \eqref{10}, so as to get the relativized counterpart of \eqref{11}, namely \eqref{17}:
\begin{align}
\label{17}
p(\lambda_{12/s}|P_{\omega_{1/s}}) \neq 0 \\ \nonumber
p(\lambda_{12/s}|P_{\omega_{2/s}}) \neq 0 \\ \nonumber
p(\lambda_{12/s}|P_{\omega_{3/s}}) \neq 0 \\ \nonumber
p(\lambda_{12/s}|P_{\omega_{4/s}}) \neq 0. \\ \nonumber
\end{align}

Now the complex system $s_{12}$ is measured. But note that, according to \textbf{Measurement Response}, the measurement system $s^*$---which is another observer with respect to $s$---will respond solely to the properties of $s_{12}$. The crucial question in a relational context becomes: The properties of $s_{12}$ \textit{relative to what other system}? To answer this question, we have to take into account two crucial facts: on the one hand, any system is on a par in RQM, and, on the other hand, physical objects do not possess inherent absolute properties \emph{per se}, independently of any observer. Hence, it is reasonable to assume that the measurement system will respond to the properties of $s_{12}$ \textit{relative to $s^*$}, represented by the relativized ontic state $\lambda_{12}/s^*$.\ In this respect, it is crucial to note that there is no guarantee that, in RQM, $\lambda_{12/s} = \lambda_{12/s^*}$. On the contrary, in general they will differ. Thus, no contradiction will arise. In effect, suppose that the relativized counterpart of \eqref{14}, that is \eqref{18} below, holds true:

\begin{align}
\label{18}
p(\lambda_{12/s^*}|P_{\omega_{1/s^*}}) = 0 \\ \nonumber
p(\lambda_{12/s^*}|P_{\omega_{2/s^*}}) = 0 \\ \nonumber
p(\lambda_{12/s^*}|P_{\omega_{3/s^*}}) = 0 \\ \nonumber
p(\lambda_{12/s^*}|P_{\omega_{4/s^*}}) = 0 \\ \nonumber
\end{align}

\noindent Still, \eqref{17} and \eqref{18} constitute no contradiction because of the difference in their relativization targets. Thus, we can safely claim that one cannot derive the conclusion of the PBR theorem in this relational context.  At this juncture, one may claim that we over-simplified the relevant details of the PBR theorem, by relativizing states of $s_1$ and $s_2$ to the \textit{same} observer.\\ 

Let us discuss another experimental situation involving two observers, say Alice and Bob, located at two different sites.\ We will show that neither in this case one can derive a relational PBR-like result.\footnote{The generalization of our conclusion to $N$ observers is trivial, and it is left as an exercise to the reader.}\ According to this scenario, Alice's task is to prepare the quantum system $s_1$ in one of the possible epistemic states $|+\rangle$ or $|\uparrow\rangle$, as required by the PBR argument (cf.\ Section \ref{2.3}). Bob has the same assignment for the quantum system $s_2$. Let us assume for the sake of argument that \textbf{Preparation Independence} holds. Then, from this experimental setting, the ontic state of $s_1$ relative to Alice, i.e.\ $\lambda_{1/A}$ should be compatible with the following probabilities:

\begin{align}
\label{19}
p(\lambda_{1/A}|P_{\uparrow/A}) \neq 0,\\ \nonumber
p(\lambda_{1/A}|P_{+/A}) \neq 0.
\end{align}

\noindent Similarly for Bob:
\begin{align}
\label{20}
p(\lambda_{2/B}|P_{\uparrow/B}) \neq 0,\\ \nonumber
p(\lambda_{2/B}|P_{+/B}) \neq 0.
\end{align}

It is worth noting that in virtue of \textbf{Preparation Independence}, Alice's preparation of $s_1$ is not influenced by (i.e.\ is independent of) Bob's preparation of the system $s_2$, and vice versa.\ Interestingly, Alice does not know how Bob prepared $s_2$. Therefore she has no information about the final state he obtained after the implementation of the preparation protocols.\ Thus, according to RQM, from Alice's perspective the system $s_2$ will have an ontic state $\lambda_{2/A}$ which differs from $\lambda_{2/B}$.\ More precisely, given the limited knowledge that Alice has about Bob's preparations---she only knows that he is preparing $s_2$ and that this system is compatible with the state $|\uparrow\rangle$ or $|+\rangle$---it follows that 

\begin{align}
\label{21}
\lambda_{2/A}=\frac{1}{\sqrt{2}}|\uparrow\rangle\otimes |up\rangle_{Bob} + \frac{1}{\sqrt{2}}|+\rangle\otimes|plus\rangle_{Bob}.
\end{align}

\noindent Hence, from her perspective the physical state of $s_2$ is represented by a superposition of the $|\uparrow\rangle$ and $|+\rangle$ states each entangled with Bob's state after the preparation, something allowed---and indeed required---by RQM, as we saw in Section \ref{2.1} while discussing the Third Person problem. Indeed, the state in \eqref{21} has notable similarities with \eqref{3}.\ As we pointed out several times already, Alice's description of $\lambda_2$ is correct even though it is not equivalent to Bob's, who does not obtain a superposition of states in virtue of his interaction with the system $s_2$ via the preparation device. An analogous argument holds for Bob, who will assign to the system $s_1$ the following ontic state:

\begin{align}
\label{22}
\lambda_{1/B}=\frac{1}{\sqrt{2}}|\uparrow\rangle\otimes |up\rangle_{Alice} + \frac{1}{\sqrt{2}}|+\rangle\otimes|plus\rangle_{Alice}
\end{align}
\noindent which is clearly different from the $\lambda_{1/A}$ obtained by Alice after her preparations. 

This fact is relevant for our discussion---and more generally for the PBR-like argument we  are exploring---since Alice and Bob have to send their prepared systems $s_1$ and $s_2$ to a measuring device $s^*$, the third observer of this scenario. As in the previous case, $s^*$ will interact only with the composite system $s_{12}$ formed  by the product of the individual states of $s_1$ and $s_2$. Remarkably, since $\lambda_{1/A}\neq\lambda_{1/B}$ and $\lambda_{2/A}\neq\lambda_{2/B}$, it follows that Alice and Bob will provide different relational descriptions of the ontic state of the complex system $s_{12}$. In effect, given \eqref{21} and \eqref{22}, we expect that $\lambda_{{12}/A}\neq\lambda_{{12}/B}$. 
In particular, for Alice the ontic state $\lambda_{{12}}$ can be represented by these two possible options: 

\begin{align}
\label{23}
\lambda_{{12}/A}=|\uparrow\rangle_A\bigg(\frac{1}{\sqrt{2}}|\uparrow\rangle\otimes |up\rangle_{Bob} + \frac{1}{\sqrt{2}}|+\rangle\otimes|plus\rangle_{Bob}\bigg) = |\omega_1\rangle_A, \\ \nonumber
\lambda_{{12}/A}=|+\rangle_A\bigg(\frac{1}{\sqrt{2}}|\uparrow\rangle\otimes |up\rangle_{Bob} + \frac{1}{\sqrt{2}}|+\rangle\otimes|plus\rangle_{Bob}\bigg) = |\omega_2\rangle_A,
\end{align}

\noindent while for Bob the ontic state $\lambda_{{12}}$ of $s_{12}$ is given by the following states:

\begin{align}
\label{24}
\lambda_{{12}/B}=|\uparrow\rangle_B\bigg(\frac{1}{\sqrt{2}}|\uparrow\rangle\otimes |up\rangle_{Alice} + \frac{1}{\sqrt{2}}|+\rangle\otimes|plus\rangle_{Alice}\bigg{)} = |\omega_3\rangle_B, \\ \nonumber
\lambda_{{12}/B}=|+\rangle_B\bigg(\frac{1}{\sqrt{2}}|\uparrow\rangle\otimes |up\rangle_{Alice} + \frac{1}{\sqrt{2}}|+\rangle\otimes|plus\rangle_{Alice}\bigg) = |\omega_4\rangle_B.
\end{align}

\noindent It is straightforward to see that these descriptions are incompatible with the products states of \eqref{10}.\ However, from Alice's perspective we have that:

\begin{align}
\label{25}
p(\lambda_{{12}/A}| P_{\omega_{1/A}}) \neq 0, \\ \nonumber
p(\lambda_{{12}/A}| P_{\omega_{2/A}}) \neq 0. \\ \nonumber
\end{align}
\noindent Once again, similarly for Bob:
\begin{align}
\label{26}
p(\lambda_{{12}/B}| P_{\omega_{3/B}}) \neq 0, \\ \nonumber
p(\lambda_{{12}/B}| P_{\omega_{4/B}}) \neq 0. \\ \nonumber
\end{align}

\noindent Interestingly, the probabilities in \eqref{25} will be zero according to Bob's perspective. By construction Bob cannot find any superposition-state for $s_2$. Similarly, the probabilities in \eqref{26} will be zero for Alice. These results simply follow from the relativization of states in RQM. Thus, there is no contradiction between Alice's and Bob's perspectives. 

Finally, if Alice and Bob send their prepared systems $s_1$ and $s_2$ to the measurement device $s^*$, the latter will interact with a composite system $s_{12/s^*}$ which will not contain any superposition of states contrary to the cases represented in \eqref{23} and \eqref{24}, so that the observer $s^*$ will not  find the complex system $s_{12/s^*}$ in any of the states $|\omega_1\rangle_A, |\omega_2\rangle_A, |\omega_3\rangle_B, |\omega_4\rangle_B$. This fact is due once again to the relativization of states, from which it follows that
\begin{align}
\label{27}
\lambda_{{12}/s^*}\neq \lambda_{{12}/A}, \\ \nonumber
\lambda_{{12}/s^*}\neq \lambda_{{12}/B}. \\ \nonumber
\end{align}

\noindent In turn, given the ontic states defined in \eqref{23}, \eqref{24}, the inequality in \eqref{27}, and from the main tenet of RQM, it follows that:
\begin{align}
\label{28}
p(\lambda_{{12}/A}| P_{\omega_{1/A}}) \neq p(\lambda_{12/s^*}| P_{\omega_{1/s^*}}), \\ \nonumber
p(\lambda_{{12}/A}| P_{\omega_{2/A}}) \neq p(\lambda_{12/s^*}| P_{\omega_{2/s^*}}), \\ \nonumber
p(\lambda_{{12}/B}| P_{\omega_{3/A}}) \neq p(\lambda_{12/s^*}| P_{\omega_{3/s^*}}), \\ \nonumber
p(\lambda_{{12}/B}| P_{\omega_{4/A}}) \neq p(\lambda_{12/s^*}| P_{\omega_{4/s^*}}). \\ \nonumber
\end{align}

Therefore, the measuring device $s^*$ will interact with a composite physical system having an ontic state completely different w.r.t.\ those described by Alice and Bob. As in the previous case, then, the relativization of states implies that the three observers involved in our scenario---namely Alice, Bob and $s^*$---will assign different ontic states to the composite system $s_{12}$, so that no contradiction can arise among these different perspectives and standard quantum predictions. The conclusion stands: RQM is neither vulnerable to the original PBR-theorem, nor to its relativized counterpart.

\section{Conclusions}
\label{conc}

In this paper we discussed throughly the relation between RQM and the PBR-theorem. This is because there is a \textit{prima-facie} significant tension between the two. RQM suggests an epistemic take on the quantum wavefunction, whereas the PBR theorem allegedly establishes that \textit{no epistemic} interpretation is empirically equivalent to standard quantum mechanics. This seems to threaten the \textit{empirical adequacy} of RQM. We argued there is no such threath. This sheds new light on both RQM and PBR. As for the former, it forces us to accept that there is an epistemic reading of the quantum wavefunction  that is both (i) empirically adequate, and (ii) is not captured by extant characterization of quantum epistemic states. As for the latter, it shows that there are limitations to the very applicability of the theorem and its conclusion that have largely gone unnoticed in the literature.\footnote{For insightful discussions on previous drafts of the paper we would like to thank [Redacted].} 

\clearpage

\bibliographystyle{apalike}
\bibliography{PhDthesis}

\begin{thebibliography}{}

\bibitem[Albert, 2013]{Albert:2013}
Albert, D.~Z. (2013).
\newblock Wave function realism.
\newblock In Ney, A. and Albert, D.~Z., editors, {\em {The Wave Function:
  Essays on the Metaphysics of Quantum Mechanics}}, pages 52--57. Oxford
  University Press.

\bibitem[Bohm, 1952]{Bohm:1952aa}
Bohm, D. (1952).
\newblock {A suggested interpretation of the quantum theory in terms of
  ``hidden'' variables. I}.
\newblock {\em Physical Review}, 85(2):166 --179.

\bibitem[Bong et~al., 2020]{Bong:2020}
Bong, K.-W., Utreras-Alarc{\'o}n, A., Ghafari, F., Liang, Y.-C., Tischler, N.,
  Cavalcanti, E., Pryde, G., and Wiseman, H. (2020).
\newblock {A strong no-go theorem on the Wigner's friend paradox}.
\newblock {\em Nature Communications}, 16:1199--2005.

\bibitem[Calosi and Mariani, 2020]{Calosi:2020}
Calosi, C. and Mariani, C. (2020).
\newblock {Quantum relational indeterminacy}.
\newblock {\em Studies in History and Philosophy of Science Part B: Studies in
  History and Philosophy of Modern Physics}, 71:158--169.

\bibitem[Candiotto, 2017]{Candiotto:2017}
Candiotto, L. (2017).
\newblock {The Reality of Relations}.
\newblock {\em {Giornale di Metafisica}}, 2:537--551.

\bibitem[Caves et~al., 2002]{Fuchs:2002}
Caves, C., Fuchs, C., and Schack, R. (2002).
\newblock {Quantum probabilities as Bayesian probabilities}.
\newblock {\em Physical Review A}, 65:022305.

\bibitem[Dorato, 2016]{Dorato:2016}
Dorato, M. (2016).
\newblock {Rovelli's Relational Quantum Mechanics, Anti-Monism, and Quantum
  Becoming}.
\newblock In Marmodoro, A. and Yates, D., editors, {\em {The Metaphysics of
  Relations}}, pages 235--262. Oxford University Press.

\bibitem[D{\"u}rr et~al., 2013]{Durr:2013aa}
D{\"u}rr, D., Goldstein, S., and Zangh{\`\i}, N. (2013).
\newblock {\em Quantum physics without quantum philosophy}.
\newblock Berlin: Springer.

\bibitem[Everett, 1957]{Everett:1957aa}
Everett, H. (1957).
\newblock {``{R}elative state'' formulation of quantum mechanics}.
\newblock {\em Reviews of Modern Physics}, 29(3):454--462.

\bibitem[Frauchiger and Renner, 2018]{Frauchiger:2018}
Frauchiger, D. and Renner, R. (2018).
\newblock {Quantum Theory Cannot Consistently Describe the Use of Itself}.
\newblock {\em Nature Communications}, 9:1--10.

\bibitem[Hance et~al., 2021]{Ladyman:2021}
Hance, J., Rarity, J., and Ladyman, J. (2021).
\newblock {Wavefucntions can Simultaneously Represent Knowledge and Reality}.
\newblock {\em arXiv:2101.06436v1}, pages 1--6.

\bibitem[Harrigan and Spekkens, 2010]{Harrigan:2010}
Harrigan, N. and Spekkens, R. (2010).
\newblock {Einstein, Incompleteness, and the Epistemic View of Quantum States}.
\newblock {\em Foundations of Physics}, 40:125--157.

\bibitem[Laudisa and Rovelli, 2019]{Laudisa:2019b}
Laudisa, F. and Rovelli, C. (2019).
\newblock {Relational Quantum Mechanics}.
\newblock {\em Stanford Encyclopedia of Philosophy}.

\bibitem[Leifer, 2014]{Leifer:2014}
Leifer, M. (2014).
\newblock {Is the quantum state real? An extended review of $\psi$-ontology
  theorems}.
\newblock {\em Quanta}, 3(1):67--155.

\bibitem[Lewis, 1986]{Lewis:1986aa}
Lewis, D. (1986).
\newblock {\em Philosophical papers}, volume~2.
\newblock Oxford: Oxford University Press.

\bibitem[Lewis et~al., 2012]{Lewis:2012}
Lewis, P., Jennings, D., Barrett, J., and Rudolph, T. (2012).
\newblock {Disticnt Quantum State Can Be Compatible with a Single State of
  Reality}.
\newblock {\em Physical Review Letters}, 109:150404.

\bibitem[Mansfield, 2014]{Mansfield:2014}
Mansfield, S. (2014).
\newblock {Reflections on the PBR Theorem: Reality Criteria \& Preparation
  Independence}.
\newblock In Coecke, B., Hasuo, I., and Panangaden, P., editors, {\em
  {Proceeding of the 11th workshop on Quantum Physics and Logic 2014}}, pages
  102--112. EPTCS 172.

\bibitem[Mansfield, 2016]{Mansfield:2016}
Mansfield, S. (2016).
\newblock {Reality of the quantum state: Towards a stronger $\psi$-ontology
  theorem}.
\newblock {\em Physical Review A}, 94:042124.

\bibitem[Nelson, 1966]{Nelson:1966aa}
Nelson, E. (1966).
\newblock Derivation of the schr{\"o}dinger equation from newtonian mechanics.
\newblock {\em Physical Review}, 150(4):1079 -- 1085.

\bibitem[Oldofredi, 2020]{Oldofredi:2020}
Oldofredi, A. (2020).
\newblock {The Bundle Theory Approach to Relational Quantum Mechanics}.
\newblock {\em Foundations of Physics}, 51(1):1--22.

\bibitem[Oldofredi and Lopez, 2020]{Oldofredi:2020b}
Oldofredi, A. and Lopez, C. (2020).
\newblock {On the Classification between $\psi-$Ontic and $\psi-$Epistemic
  Ontological Models}.
\newblock {\em Foundations of Physics}, 50(11):1315--1345.

\bibitem[Pusey et~al., 2012]{PBR:2012}
Pusey, M.~F., Barrett, J., and Rudolph, T. (2012).
\newblock On the reality of the quantum state.
\newblock {\em Nature Physics}, 6(8):475--478.

\bibitem[Reich, 2011]{Reich:2011}
Reich, E. (2011).
\newblock {Quantum theorem shakes foundations}.
\newblock {\em Nature}.

\bibitem[Rovelli, 1996]{Rovelli:1996}
Rovelli, C. (1996).
\newblock {Relational Quantum Mechanics}.
\newblock {\em International Journal of Theoretical Physics}, 35(8):1637--1678.

\bibitem[Rovelli, 2016]{Rovelli:2016}
Rovelli, C. (2016).
\newblock {An Argument Against the Realistic Interpretation of the Wave
  Function}.
\newblock {\em Foundations of Physics}, 46:1229 -- 1237.

\bibitem[Rovelli, 2018]{Rovelli:2018}
Rovelli, C. (2018).
\newblock {Space is blue and birds fly through it}.
\newblock {\em Philosophical Transactions of the Royal Society A, Physical and
  Engineering Sciences}, 376(2123):2017.0312.

\bibitem[Schlossauer and Fine, 2012]{Schlosshauer:2012}
Schlossauer, M. and Fine, A. (2012).
\newblock {Implications of the Pusey-Barrett-Rudolph Quantum No-Go Theorem}.
\newblock {\em Physical Review Letters}, 108:260404.

\bibitem[Schlossauer and Fine, 2013]{Schlosshauer:2013}
Schlossauer, M. and Fine, A. (2013).
\newblock {Is the Pusey-Barrett-Rudolph Theorem Compatble with Quantum
  Nonseparability?}
\newblock {\em arXiv:1306.5805v1}, pages 1--4.

\bibitem[Schlossauer and Fine, 2014]{Schlosshauer:2014}
Schlossauer, M. and Fine, A. (2014).
\newblock {No-go theorem for the composition of quantum system}.
\newblock {\em Physical Review Letters}, 112(7):070407.

\bibitem[Smerlak and Rovelli, 2007]{Smerlak:2007}
Smerlak, M. and Rovelli, C. (2007).
\newblock {Relational EPR}.
\newblock {\em Foundations of Physics}, 37:427--445.

\bibitem[Wallace, 2012]{Wallace:2012aa}
Wallace, D. (2012).
\newblock {\em The emergent multiverse. Quantum theory according to the
  {E}verett interpretation}.
\newblock Oxford: Oxford University Press.

\end{thebibliography}
\end{document}